\documentstyle[11pt,newpasp,twoside,epsf]{article}
\reversemarginpar

\begin{document}
\title{Timing of the young pulsar J1907+0918}
\author{K.~Xilouris}
\affil{Center Development Lab, NRAO, Charlottesville, VA 22903, USA}
\author{D.R.~Lorimer}
\affil{Arecibo Observatory, HC3 Box 53995, Arecibo, PR 00612, USA}

\begin{abstract}
We have extensively searched for periodic signals in the area of the
soft gamma-ray repeater SGR~1900+14, at 430 and 1410 MHz with the
Arecibo telescope. Our observations did not reveal the 5.16-s
periodicity reported for the magnetar down to a 430-MHz flux density
limit of 150 $\mu$Jy. During the search we discovered a 226-ms radio
pulsar, PSR~J1907+0918. The period derivative implies the
characteristic age for this pulsar is only 38 kyr. Independent
evidence in support of this apparent youth are the unusually high
degree of circular polarization and its relatively flat radio
spectrum. The close proximity of PSR~J1907+0918 to SGR~1900+14
suggests that they may have evolved from a massive binary system.
Regardless of this possibility, it is presently not clear to
us that SGR~1900+14 is associated with the supernova remnant G42.8+0.6.
\end{abstract}

\section{The search for radio pulsations from SGR~1900+14}

In early June 1998 we observed the soft gamma-ray repeater SGR~1900+14
using the Arecibo telescope seven days after the source became active
following a long period of quiescence. The search for radio pulsations
at 430 MHz and 1.4 GHz was carried out with the Penn State Pulsar
Machine (PSPM), a filterbank which records the total power outputs of
the receiver over $128\times60$ kHz frequency channels every 80$\mu$s.

Our search did not reveal the 5.16-s period reported for SGR~1900+14
by Kouveliotou et al.~(1998). Based on our observations we place an
upper limit of approximately 150 $\mu$Jy to the flux density of the
magnetar at frequencies around 430 MHz.  Following the announcement of
a low-frequency detection of this pulsar by Shitov (see contribution
elsewhere in these proceedings) we observed the magnetar using the 47
MHz dipole feed. Although this system could detect B0950+08 and
B0823+26 we were unable to detect the magnetar.  Further, more
sensitive, low-frequency Arecibo observations would be worthwhile.

The 1410-MHz observations did, however, reveal the presence of a very
promising 113-ms pulsar candidate with a dispersion measure of 350
cm$^{-3}$ pc. Subsequent observations made around the end of September
both at Arecibo and Effelsberg, confirmed the existence of the pulsar
(PSR~J1907+0918) and identified its true period to be 226 ms (Xilouris
et al.~1998, IAUC No.~7023).

\section{Timing observations of PSR~1907+0918}

Follow-up timing results show that PSR~J1907+0918 is an interesting
radio pulsar in its own right.  Regular timing observations using the
PSPM were initiated in mid October 1998. A standard {\sc tempo}
analysis of pulse time-of-arrival measurements spanning a 9-month
baseline yields the following timing solution: R.A. (J2000) 19 h 07 m
22.4 sec, Dec. 09$^{\circ}$ 18' 31.8'', $P=0.226106270831$ sec,
$\dot{P} = 94.286 \times 10^{-15}$, these parameters apply to the
reference MJD 51216.  Uncertainties for each parameter are one unit of
the least significant digit quoted. Current post-fit residuals are
98.6 $\mu$s. In spite of present covariances between position and
$\dot{P}$, it is clear from high-precision period measurements over
the 9-month baseline that the quoted $\dot{P}$ is correct. The
characteristic age is 38 kyr and implied dipole surface magnetic field
is $4.7\times10^{12}$ G.

\section{Discussion}

Apart from globular cluster pulsars, PSR~J1907+0918 and SGR~1900+14
are the closest pair of neutron stars in the sky that do not presently
consitute a binary. The angular separation between them is $\sim$ 2
arcmin.  An assumed distance of 5 kpc implies a spatial separation
of 3.2 pc, while a distance of 7 kpc implies a separation of 4.5
pc.  Either this close proximity is simply a coincidence, or the
neutron stars both originated from a disrupted massive binary system.
Such scenarios have been invoked to explain the proximity of the Crab
pulsar to B0525+21 (Gott, Gunn \& Ostriker, 1970, ApJ {160}, L91)
and PSR~B1853+01 and PSR~B1854+00 (Wolzczan Cordes \& Dewey 1991, 
ApJ, {372}, L99).

Regardless of the fact that these two neutron stars may have had a
common origin, we would like to point out that it is presently not
clear to us whether the nearby supernova remnant G42.8+0.6 is
associated with SGR~1900+14 (which has an estimated age of 10 kyr;
Kouveliotou et al.~1999, ApJ, {510} L115) or PSR~J1907+0918 (with
a characteristic age of 38 kyr). As mentioned above, Shitov
has recently detected 5.16-s pulsations from
SGR~1900+14 at 100 MHz and determined a dispersion measure of 281.4 cm$^{-3}$
pc. Based on this and the dispersion measure for J1907+0918, both
these neutron stars are at a comparable distance from the Earth (5--7
kpc). If PSR~J1907+0918 is associated with G42.8+0.6 then the spatial
separation between them is 20pc (assuming a distance to the remnant of
5 kpc) or 28pc (assuming a distance to the remnant of 7 kpc). The
transverse velocity required for the remnant and the pulsar to be
associated is then between 550--760 km s$^{-1}$.

It should also be noted that, since this region of the Galactic plane
has a high density of supernova remnants and pulsars, it is possible
that neither PSR~J1907+0918 nor SGR~1900+14 have any connection with
G42.8+0.6. Future VLBI proper motion measurements of PSR~J1907+0918,
perhaps using Arecibo-Effelsberg-GBT, would certainly help to
clarify this situation.

\acknowledgments
We wish to thank A. Wolszczan and D. Backer for providing access to their
datataking equipment and hence making these observations possible.
We would also like to thank F. Camilo and I. Stairs for useful discussions
concerning the Arecibo timing observations. Arecibo Observatory is 
run by Cornell University under contract with the National Science Foundation.
\end{document}